\documentclass[aps,prb,twocolumn,showpacs]{revtex4}
\usepackage{graphics}
\usepackage{subfigure}
\usepackage{epsfig}
\usepackage{epsf,epic}
\usepackage{color}
\usepackage{subfigure}
\usepackage{amsmath}
\usepackage{amssymb}
\usepackage{amsfonts}
\usepackage{wrapfig}
\usepackage{pstricks}
\usepackage{multirow}
\usepackage{bm}
\usepackage{dcolumn}
\newcommand{\etal}{\textit{et al.\ }}

\begin{document}
\title{Raman study of the vibrational modes in ZnGeN$_2$ (0001)}
\author{Eric W. Blanton, Mark Hagemann, Keliang He,\footnote{Present address: Advanced Technology Development, GlobalFoundries, 2070 Route 52​,​
Hopewell Junction, NY 12533} Jie Shan,\footnote{Present address: Department of Physics and Center for Two-Dimensional and Layered Materials, The Pennsylvania State University, University Park, Pennsylvania 16802-6300, USA} Walter R. L. Lambrecht, and Kathleen Kash}
\affiliation{Department of Physics, Case Western Reserve University, 10900 Euclid Avenue, Cleveland, OH-44106-7079}
\begin{abstract}
A Raman spectroscopy study was carried out for ZnGeN$_2$ with direction of propagation along the (0001) crystallographic direction on hexagonal single crystal platelets obtained by reaction of gaseous ammonia with a Zn-Ge-Sn liquid alloy at 758 $^\circ$C. The sample geometry allowed measurement of the $a_2$ and  $a_1$ Raman modes. First-principles calculations were carried out of the spectra. Measurements with crossed polarizers yielded spectra that agreed well with first-principles calculations of the $a_2$ modes.
Measurements with parallel polarizers should in principle provide the
$a_{1L}$ modes.  However, for most of the Raman modes, the LO-TO 
splitting was calculated to be very small, and for  the few modes which were predicted
to have larger LO-TO splittings, the LO mode was not observed. This absence 
is tentatively explained in terms of overdamped LO-plasmon coupling.  LO-TO mode crossing
was identified by comparing the calculated eigenvectors and intensities of individual modes. Some features in the experimental spectra were identified as arising from critical points in the phonon density of states and are indicative of the degree of disorder on the cation lattice. 
\end{abstract}
\maketitle
\section{Introduction}
ZnGeN$_2$ is a heterovalent ternary analogue of wurtzite GaN, obtained 
conceptually by substitution of each pair of Ga atoms, belonging to group III, 
by a Zn (group II) and Ge (group IV) atom. There has been increasing interest lately in the heterovalent ternary nitrides. This interest is motivated in part by the search for alternatives to the binary nitrides that would be composed entirely of abundant elements. In addition, the increased complexity of the compounds offers opportunities for developing doping strategies, engineering defects, and tailoring properties, that are not available in the simpler, binary nitrides. The prospect of combining the two families of materials in heterostructures offers additional opportunities. For example, while the band gap of ZnGeN$_2$ is within 100 meV or so of that of GaN, the band offset is predicted to to be as large as 40 percent of the band gap.\cite{Punya13} This situation offers the prospect of designing novel nitride heterostructures based upon the type-II character of the interface, taking advantage also of the close lattice match and similar optimal growth temperatures of the two materials.\cite{Han} 

The increased complexity of the lattice, compared to the binary nitrides, 
also presents a possibly more difficult task when attempting to determine 
fundamental properties of the materials. 
Depending on the growth conditions, ZnGeN$_2$ may exhibit either a disordered (wurtzite-like)  
or ordered structure.\cite{Blanton13} 
In a recent paper, it was shown that even 
the disordered phase may obey locally the octet rule of having exactly two Zn atoms 
and two Ge atoms as nearest neighbors of each N atom.\cite{Quayle15} This phase
involves a disordered mixture at the atomic scale of two stacking arrangements
of rows of atoms in the basal plane, each row containing alternating Zn 
and Ge atoms, corresponding to two simple ordering schemes with 
space groups Pna2$_1$ and Pmc2$_1$. To date, 
only the Pna2$_1$ structure, which has a calculated energy of formation 120 meV/formula unit less than that of the  Pmc2$_1$ structure,\cite{Quayle15}  has been observed. 

The Pna2$_1$ structure of ZnGeN$_2$ was observed first by neutron diffraction.\cite{Wintenberger} Because it has 16 atoms per unit cell, its vibrational spectrum is much more complex than that of the wurtzite structure. The vibrational spectrum of ZnGeN$_2$  was studied computationally 
in a series of papers.\cite{Lambrechtzgn05,Paudel08,Lambrechtbook} The first report of a measured Raman spectrum was for polycrystalline material. The unpolarized spectrum showed no well-resolved Raman peaks.\cite{Viennois} Subsequently, Raman spectra were measured for hexagonally faceted single crystals of diameters of a few microns and lengths of tens of microns along the c axis
by Peshek \etal  \cite{Peshek08} Using this geometry and exploiting 
Raman polarization dependent selection rules, it was possible to measure 
the $a_1$ and $b_2$ TO modes. In addition, there was clear correlation between some features in the measured spectra with peaks in the calculated phonon density of states. 

In this paper we report a new growth 
procedure, which leads to platelet shaped crystallites with the 
plane of the platelet being the  $c$-plane. 
This geometry offers the opportunity to measure the $a_2$ Raman spectrum and, in principle,  the $a_1$ LO modes, and is the preferred growth direction for heterostructures and films grown on substrates. Since the same orientation 
will occur in films grown on the basal plane, it is important to 
establish the corresponding Raman spectrum experimentally as it 
could become useful to monitor the quality of crystal growth. 
By using cross-polarized as well as parallel-polarized
spectra for different orientations relative to the hexagonally shaped plates, 
we were able to identify the $a_2$ modes and the $a_1$ modes. 
Good correspondence between theory and experiment was obtained for the $a_2$ 
modes and the $a_1$ modes that show weak LO-TO splitting. 
However, for the few modes that 
show a larger LO-TO splitting, the LO peaks were not visible in the experimental
spectra. This result is explained in terms of overdamped LO-plasmon coupling. 
While one can not in principle associate the TO modes to corresponding LO modes on a one-to-one basis, an approximate correspondence can be 
established based on their Raman intensities, which indicates that these modes  have similar eigenvectors. Within this correspondence, we show computationally 
that a mode crossing occurs. 
The modes which are found to be strongly Raman active are found to be only 
weakly infrared active, and vice versa. 

\section{Methods}

\subsection{Experimental}

ZnGeN$_2$ was synthesized in a quartz tube furnace by exposing a Ge-Zn-Sn liquid to gaseous ammonia at 758 ~$^\circ$C.  
To form the liquid, a small amount of Sn (50 mg) was placed on a [111] oriented Ge wafer.  A small amount of Ge melted to form a liquid alloy that was in equilibrium with the underlying solid Ge.  
The Zn was supplied by a heated Zn crucible upstream of the Ge wafer, which maintained a Zn pressure in the growth chamber
of approximately 0.03 atm.  Based on equilibrium data, the predicted liquid composition was 31\% Sn, 15\% Zn, and 54\% Ge.
The NH$_3$ pressure was maintained at 0.31 atm and the H$_2$ carrier gas pressure was 0.63 atm.
The growth time was 4.0 hours.
Platelet-shaped crystals approximately 20 microns in diameter formed on the surface of the melt.
The resulting ZnGeN$_2$ platelets had ordered cation lattices, 
corresponding to the Pna2$_1$ spacegroup as evidenced by x-ray diffraction.  The lattice parameters (a=$6.44\pm0.01$ \AA, b=$5.467\pm0.006$ \AA, c=$5.191\pm0.004$ \AA)
indicate the orthorhombic distortion associated with ordering.

Micro-Raman measurements were performed using a 633 nm laser focused through a 50$\times$ objective to a spot size of approximately 1 micron. The scattered light was collected in reflection using the same objective. The sample was rotated in order to change the direction of incident polarization with respect to the crystallographic axes. 

\subsection{Computational}
The vibrational modes were calculated using density functional perturbation theory,\cite{Gonze1,Gonze2} also called linear response theory, within a plane wave basis set pseudopotential method using the ABINIT code.\cite{abinit}
The calculations were performed in the local density approximation (LDA),\cite{PerdewWang92} and used the relativistic Hartwigsen, Goedecker, Hutter (HGH) 
norm-conserving pseudopotentials.\cite{Hartwigsen98} A large plane wave cut-off
of 80 Hartree was used and the Brillouin zone integration used a $4\times4\times4$ integration mesh. For Zn the $3d$ electrons were treated as valence electrons while 
for Ge they were included in the pseudized core. 

\section{Results}

\begin{figure}
\includegraphics[width=7cm]{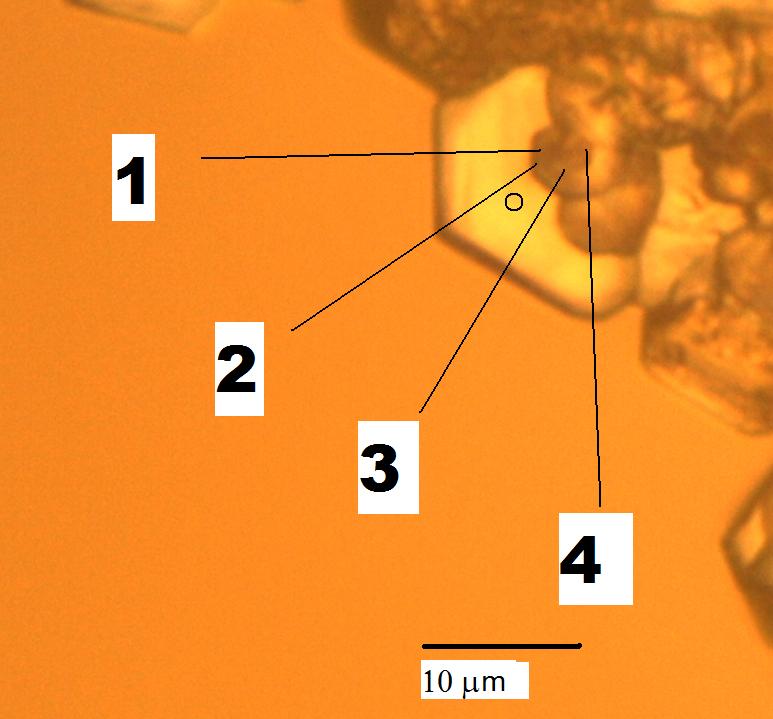}
\caption{(Color online) Hexagonal platelet of ZnGeN$_2$ indicating the excitation spot size 
and location, and 
directions of polarizations for the Raman measurements.  Using the analysis outlined in the text, we determined 
that the $x$ and $y$ axes of the crystal correspond to directions 4 and 1, 
respectively. \label{figplatelet}}
\end{figure}

The Raman spectra were recorded for a hexagonal platelet for several polarization
directions as indicated in Fig. \ref{figplatelet}.
The scattering geometry has the wavevector of the laser light at nearly 
normal incidence to the plane of the platelets; that is, along the $c$-direction. 
When the incoming and scattered light have  parallel polarizations, the 
$a_1$ symmetry modes are excited. The Raman tensor for $a_1$ symmetry 
has different components for the $xx$, $yy$ and $zz$ polarizations of the 
incoming and scattered light.  Since the wavevector is along $z$  and  a vector's 
$z$ component belongs to the $a_1$ irreducible representation of the 
$C_{2v}$ point group, the spectrum measured in this way should 
correspond to the longitudinal optical modes.  Although the $a_1$ modes 
contain displacements of the atoms along $x$ and $y$ as well as $z$, only their 
$z$-components contribute to the Raman tensor. For example, 
the Raman tensor element
\begin{equation}   
R_{xx}\propto \sum_{i\alpha} 
\frac{\partial \chi_{xx}}{\partial u_{i\alpha}}u_{i\alpha}^m
\end{equation}
involves the displacements eigenvector $u_{i\alpha}^m$ of the $m$-th mode
of $a_1$ symmetry, with $\alpha$ the Cartesian component and $i$ the index 
of the atom in the unit cell. Although there is a sum over $i$ and $\alpha$, 
because of the third rank tensorial 
nature of the derivative of the susceptibility versus atomic displacements  
only the $xxz$ elements of the tensor are non-zero and thus only the $z$ 
displacements contribute.  
On the other hand, the $a_2$ modes correspond to the $R_{xy}$ Raman tensor
and should be measurable with crossed polarizers. However, we do not have an independent 
measure of which is the $x$ and which is the $y$ direction on the platelet.
We assume that $x$ must point either to the corner of the hexagonal plate
or to the middle of the side, but there are still multiple corners and sides that could be the crystal's $x$ or $y$ directions.  
We utilized the selection rules to narrow down the choices.
For a general direction in the plane, and parallel 
polarizers, the intensity should be proportional to 
\begin{equation} 
I\propto |\cos^2{\phi}R_{xx}+\sin^2{\phi}R_{yy}+2 \sin{\phi}\cos{\phi}R_{xy}|^2
\end{equation}
with $\phi$ the azumuthal angle measured from the $x$ axis. For parallel polarizations, if $\phi$ is in a direction between $x$ and $y$
both $a_1$ and $a_2$ modes will be present in the spectrum.  Only when $\phi=0$ or $\phi=90^\circ$ does the spectrum contain no $a_2$ modes.  

\begin{figure}
\includegraphics[width=9cm]{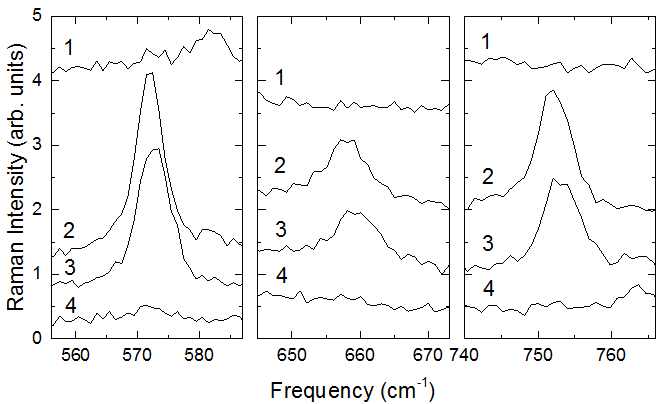} 
\caption{Raman spectra for parallel polarizations of incoming 
and scattered light for the four different directions shown in 
Fig. \ref{figplatelet} in the frequency 
ranges of the main $a_2$ peaks.  The $a_2$ peaks are suppressed in directions 1 and 4, which
indicates that one of these two directions is $x$ and one is $y$.\label{fig4dir}}
\end{figure}

To determine which are the $x$ and $y$ axes, we first show the 
spectra under parallel polarizers in Fig. \ref{fig4dir}
for the four directions shown in Fig. \ref{figplatelet}. 
The peaks at 572, 658, and 753 cm$^{-1}$
are the three strongest measured $a_2$ modes.
These modes can be identified 
with the calculated $a_2$ modes at 568, 654 and 754 cm$^{-1}$ reported in Ref. \onlinecite{Lambrechtzgn05}.  Their Raman intensities 
were reported in Ref. \onlinecite{Paudel08}. In the present calculation, these modes occur at 577, 664 and 764 cm$^{-1}$.

\begin{table}
\caption{Calculated and measured $a_2$ mode frequencies in cm$^{-1}$
and corresponding calculated Raman tensor (in atomic units). \label{tabmodea2}}
\begin{ruledtabular}
\begin{tabular}{ccddcd}
Expt. & Calc.  & \multicolumn{1}{c}{Diff.} & \multicolumn{1}{c}{$R_{xy}$ (10$^{-4}a_0^{-1}$)} &Other\footnote{Calculations from Lambrecht {\em et al.}  \cite{Lambrechtzgn05}} & \multicolumn{1}{c}{Diff.}  \\  \hline
133.9 & 132.3 & -1.6 & 3.37 & 129.9 & -4.0 \\
160.0 & 163.5 & 3.5  & 1.57 & 163.0 & 0 \\
186.0 & 184.5 & -1.5 & 3.05 & 182.3 & -2.2 \\
202.2 & 202.1 & -0.1 & 6.20 & 200.6 & -1.6\\
265.4 & 269.4 & 4.0  & 5.27 & 266.5 & 1.1 \\
334.1 & 334.6 & 0.5  & 8.12 &340.7 & 6.6 \\
470.0 & 472.1 & 2.1  & 1.02 & 470.8 & 0.8 \\
565.4 & 553.1 & -12.3& 8.44 & 555.2 & -10.2\\
572.4 & 577.4 & 5.0  & 17.64& 567.6 & -4.8 \\
658.1 & 664.5 & 6.4  & 10.58& 653.6 & -4.5 \\
753.0 & 764.0 & 11.5 & 20.53 & 753.6 & 0.6 \\
785.9 & 798.5 & 12.6 & 3.82 & 819.9 & 34 \\
Max diff &   & 12.6 &     &  & 34 \\
RMS diff &   & 6.7  &     &  & 10.8 \\
\end{tabular}
\end{ruledtabular}
\end{table}

According to Equation 2,  when $\phi$ is in the $x$ or $y$ direction, no $a_2$ modes should be present.
Clearly for directions 1 and 4 the modes at 572, 658 and 753 cm$^{-1}$ are almost completely suppressed, and thus we identify one of the directions 1 or 4 as 
the $x$ direction  and the other as the $y$ direction.  Which is which 
will be determined below on the basis of the $a_1$ modes. 

\begin{figure}
\includegraphics[width=9cm]{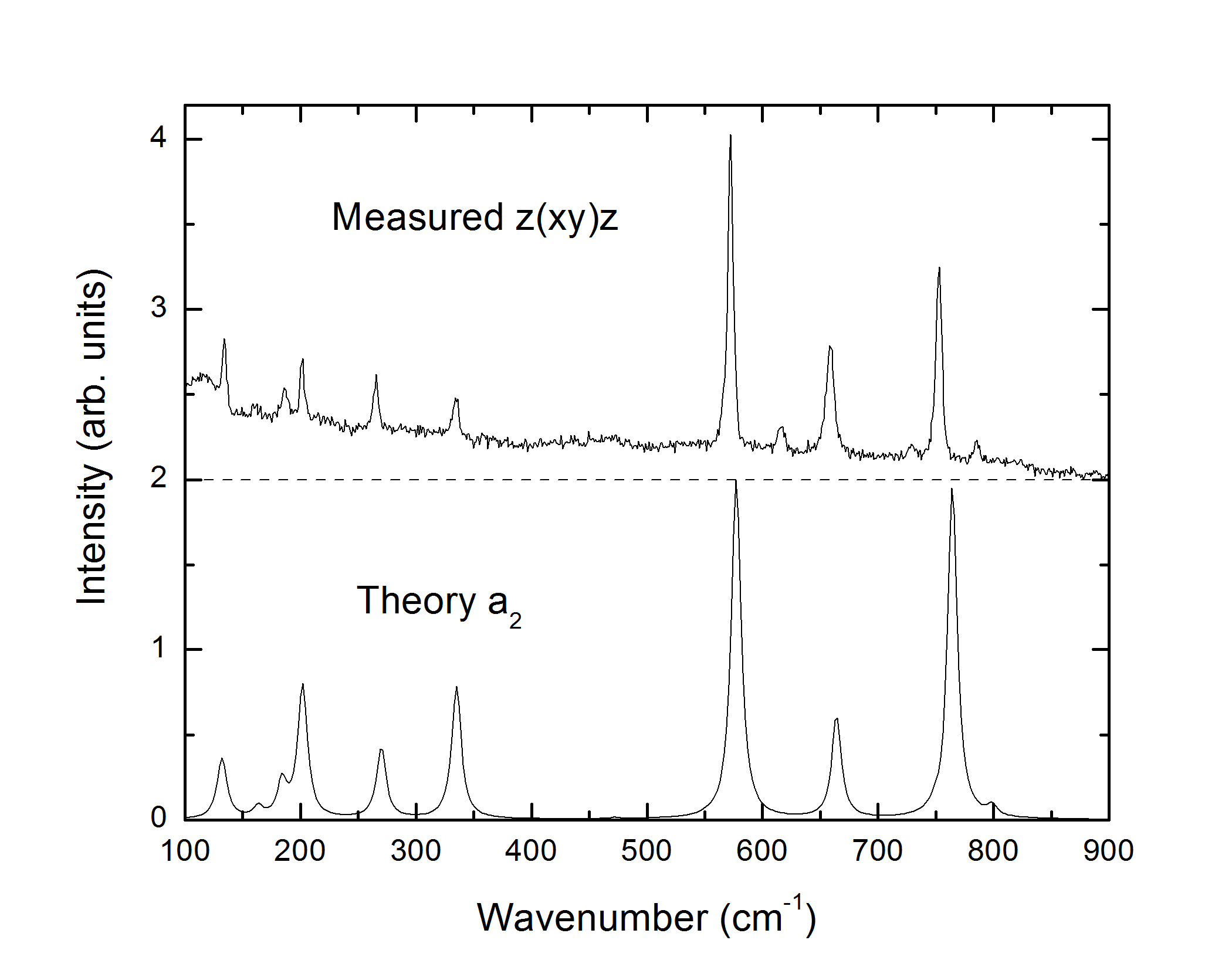}
\caption{Raman spectrum under crossed polarizers, showing the $a_2$ modes and compared with the calculated spectrum.\label{figcrossed}}
\end{figure}

The spectrum under crossed polarizers with incoming polarization along 
one of these directions is shown in Fig. \ref{figcrossed}. We can now clearly 
identify all twelve  $a_2$ modes. The measured values are compared with the 
present calculation and a previous one\cite{Lambrechtzgn05}  
in Table \ref{tabmodea2}, and the differences between these and the experimental values is noted.  The new calculation using a more accurate
pseudopotential gives a slightly smaller maximum and average difference. 
Most frequencies are obtained to within 1\% and the maximum difference between experimental and calculated values is 2\%. 
The small peak at 617 cm$^{-1}$ 
may result from incomplete suppression of the strong $a_1$ peak by the crossed polarizers, but the small peak at 726 cm$^{-1}$ cannot 
be accounted for in this manner and is unexplained. 

\begin{figure}
\includegraphics[width=9cm]{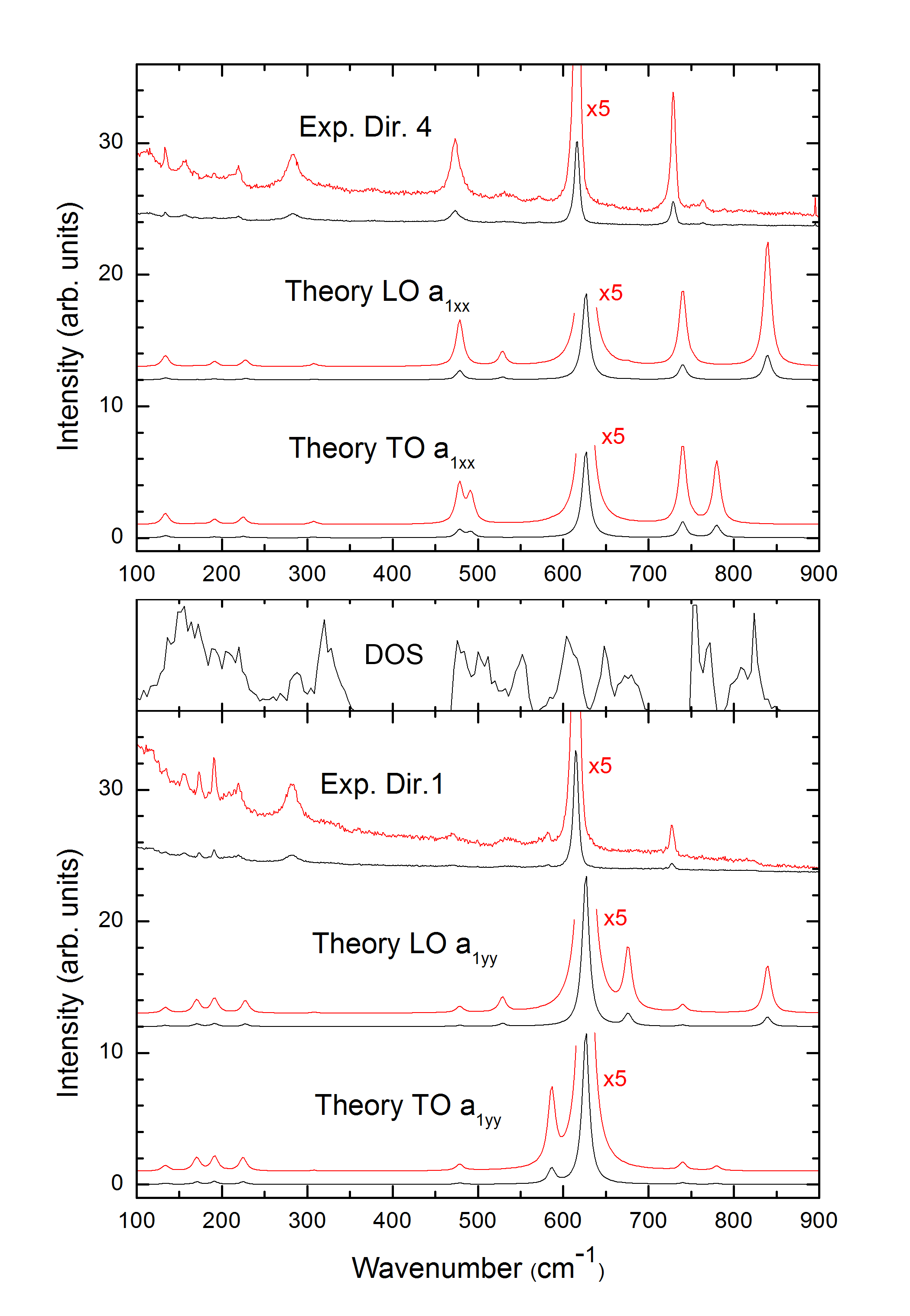}
\caption{(Color online) Raman spectra for parallel polarizers for directions 1 and 
4, compared to the computed $a_1$ spectra for, respectively, the $yy$ and $xx$ components  for both TO and LO modes. The middle panel 
shows the phonon density of states. 
\label{figpar}}
\end{figure}

In Fig. \ref{figpar} we show the spectra under parallel polarizers 
for directions 1 and 4 compared with calculated Raman spectra 
for the $yy$ and $xx$  Raman tensor components, respectively, 
for the longitudinal and transverse $a_{1}$ modes. 
 First we note that 
the spectrum for direction 1 best corresponds to the calculated 
spectrum for the $a_{1yy}$  Raman tensor  and for direction 4 to the 
$a_{1xx}$ Raman tensor.  The collection times for the measured spectra 
in directions 1 and 4 
were identical and the spectra were scaled by the same factor in the figure.  
The four calculated spectra were 
 scaled by a common factor in order to make the measured and calculated intensities roughly comparable. We can see that the absolute intensity as well as the 
ratio of the strongest peak (at 614.5 cm$^{-1}$) to the other peaks is higher for the experimental direction 1 and for the calculate $yy$ component.
Secondly, 
the details of the spectrum in the low frequency range also match best when 
identifying direction 4 with $x$ and direction 1 with $y$.  
In the experimental spectra, the peak at 132.8 cm$^{-1}$ is stronger in direction 4, and the peaks at 173.0 and 191.4 cm$^{-1}$
are stronger in direction 1.  These peaks correspond to the calculated $a_1$ modes at 133.9, 170.5, and 191.7 cm$^{-1}$.
The calculated peak positions for $a_1$ are listed in Table \ref{taba1}.
 The calculated peak at 133.9 cm$^{-1}$ is predicted to be stronger in the $x$ direction, while the peaks at 170.5 and 191.7
 cm$^{-1}$ are predicted to be stronger in the $y$ direction.

Although the material is mostly ordered as evidenced by the lattice parameters, there is possibly some disorder on the lattice 
and thus we can expect that phonon density of states (DOS) features would appear in the measured spectra.  Generally, 
the presence of DOS features in the Raman spectra of the present material is much less pronounced than in the  
spectra of the material grown by  Peshek \etal  \cite{Peshek08}.  However, there are some features in the measured spectra 
which we believe might be associated with DOS features because of their proximity to predicted DOS features and because of the lack of variablity 
of the intensities and shapes of the features with polarization direction.
A comparison of the measured spectra with  the calculated DOS\cite{Peshek08,Paudel08}
is shown in Fig. \ref{figpar}.
There is a feature in the measured spectra at 156.7 cm$^{-1}$ that has the same intensity for both directions of polarization.  This feature is most
 likely associated with a peak in the DOS.  Likewise, the feature at 219.4 cm$^{-1}$ is the same for both directions and coincides with a 
steep ledge in the DOS.  Theory also predicts a DOS peak at 225.2 cm$^{-1}$ that is very close to the measured feature.  There is a large broad measured feature at 283.4 cm$^{-1}$ that is the same for 
both directions of polarization.  The small predicted peak at 307.5 cm$^{-1}$ might be contributing to this feature as well.  

There is a large feature at 473 cm$^{-1}$ that is much stronger in the direction 4 spectrum than in direction 1.  This feature 
 corresponds to the predicted peak at 478.5 cm$^{-1}$, which is indeed 
predicted to be more intense in the $x$ direction.  The dominant 
feature in all spectra is the intense peak measured at 614 cm$^{-1}$, which corresponds to the 626.6 cm$^{-1}$ peak 
in the calculated spectra.
  This peak is observed to be more intense in direction 1 and is predicted to be more intense in the $y$ direction.  Another intense feature is the peak at
 728.7 cm$^{-1}$ in the measured spectra, which corresponds to the predicted peak at 740.5 cm$^{-1}$.
With all three of these peaks, the relative intensities in each direction of polarization are consistent with our assignment of direction 4 being the $x$ direction
 and direction 1 being the $y$ direction.

\begin{table*}
\caption{Measured and calculated TO and LO $a_1$ mode frequencies in cm$^{-1}$, Raman tensor elements for $xx$ and $yy$ polarizations and for TO and LO modes, $R_{xx}^{T(L)}$, $R_{yy}^{T(L)}$ (in atomic units) 
and IR oscillator strengths, $S_z$ (atomic units).\label{taba1}}
\begin{ruledtabular}
\begin{tabular}{cccddddc}
Expt.& \multicolumn{7}{c}{Calculation} \\
     & $\omega(a_{1T})$ & $\omega(a_{1L})$ & \multicolumn{1}{c}{$R_{xx}^T$} & \multicolumn{1}{c}{$R_{yy}^T$} & \multicolumn{1}{c}{$R_{xx}^L$} & \multicolumn{1}{c}{$R_{yy}^L$} & $S_z$ \\
$cm^{-1}$ & $cm^{-1}$ & $cm^{-1}$ & \multicolumn{1}{c}{$10^{-4} a_0^{-1}$} &  \multicolumn{1}{c}{$10^{-4}a_0^{-1}$} &  \multicolumn{1}{c}{$10^{-4}a_0^{-1}$} &  \multicolumn{1}{c}{$10^{-4}a_0^{-1}$} & 
\multicolumn{1}{c}{a.u.=253.264 m$^2$/s$^2$} \\ \hline
132.8 & 133.9 & 133.9 & 4.3 & 3.1 & 4.3 & 3.1 & $6.3\times10^{-9}$ \\
173.0 & 170.5 & 170.6 & 1.2 & 5.3 & 1.3 & 5.3 & $4.8\times10^{-7}$ \\
191.4 & 191.7 & 191.7 & 3.5 & 5.9 & 3.5 & 5.9 & $2.0\times10^{-7}$ \\
219.4 & 225.2& 227.9 & 4.6 & 6.2 & 4.4 & 6.0 & $2.6\times10^{-5}$ \\
283.4 & 306.8& 307.5 & 3.6 & 1.9 & 3.4 & 1.8 & $9.9\times10^{-6}$ \\
473.0 & 478.5 & 478.6 &16.0 & 6.5 & 17.2& 6.4 & $1.7\times10^{-5}$ \\
536.0 & 491.4 & 528.6 &13.8 & 8.0 & 10.0& 10.3& $1.1\times10^{-3}$ \\
580.0 & 586.5 & 676.2 & 3.0 & 24.5 & 3.6 & 24.0& $1.0\times10^{-3}$ \\
614.5 & 626.6 & 626.6 &61.7 & 81.6& 61.7& 81.6& $1.8\times10^{-9}$ \\
728.7 & 740.5& 740.5 &29.0 & 9.0 & 28.6 & 8.7& $6.2\times10^{-7}$ \\
760.0 & 779.4 & 839.0 & 26.5 & 7.2& 39.5 &24.3& $5.4\times10^{-4}$ \\
\end{tabular}
\end{ruledtabular}
\end{table*}

In order to better understand
 the relations between the TO and LO modes we first examine the theoretical results. 
The TO and LO modes are listed in Table \ref{taba1} along with some of their 
associated calculated quantities.  For completeness' sake we also report the 
calculated $b_1$ and $b_2$ modes in the appendix, although it was 
not possible to measure these with the present samples and scattering geometry.
In the table we have assigned each LO mode to a corresponding TO mode.  
Strictly speaking, one cannot make a one-to-one correspondence
between TO and LO modes of the same symmetry. The LO modes result 
from diagonalizing a force constant matrix, which includes 
long-range forces resulting from the coupling of the Born effective charges
to the electric field produced by the dipoles for longitudinal 
modes in the limit of the wavevector ${\bf q}\rightarrow 0$. These forces are
absent for transverse modes. In the calculations, this long-range 
electric field (here along $z$) 
is modeled as a static electric field, and we obtain 
a different set of modes when this field is included, compared
to when it is not included.  This result is referred
to as the ``non-analyticity'' in the $z$-direction. Since the 
frequencies with (LO) and without (TO) 
this non-analyticity result from diagonalizing different  $12\times12$ matrices (including one zero eigenvalue) in the two cases, 
there is no reason why the 
eigenvalues and eigenvectors 
should match up on a one-to-one basis. Nonetheless, there is a significant 
similarity in eigenvectors between most mode pairs and hence, in an approximate way, 
one could identify the modes according to the overlap between the LO and TO 
eigenvectors rather than simply ordering them according to increasing 
frequency. This similarity in their eigenvectors is manifested in the 
Raman tensor values, and the LO and TO mode correspondences were assigned with these
considerations in mind.

The chosen LO-TO mode correspondences also become apparent when considering the calculated oscillator strengths of
 the modes, listed in Table \ref{taba1}.  The oscillator strength is proportional to the long-range electric 
field set up in the crystal as a result of the normal mode oscillation and so should be related to the size of the LO-TO shift.
Three of the eleven modes have significantly larger calculated oscillator strengths 
than the other modes and accordingly the LO-TO shifts associated with these assigned mode pairs are the largest of the 
eleven mode pairs.  We note that in crystals with inversion symmetry modes are 
either Raman active or IR active but not both. Although in the present 
case we do not have inversion symmetry, there still appears to be clear distinction 
between modes that are strong in IR and weak in Raman or vice versa. 
Calculated IR spectra were reported in Ref. \onlinecite{Paudel08}, Fig. 1, 
and show indeed only three strong peaks in the 
${\rm Im}(\varepsilon)$ spectrum, corresponding to TO modes,  
and three corresponding peaks in the $-{\rm Im}\varepsilon^{-1}$ 
spectra corresponding to LO modes. 

\begin{figure}
\includegraphics[width=9cm]{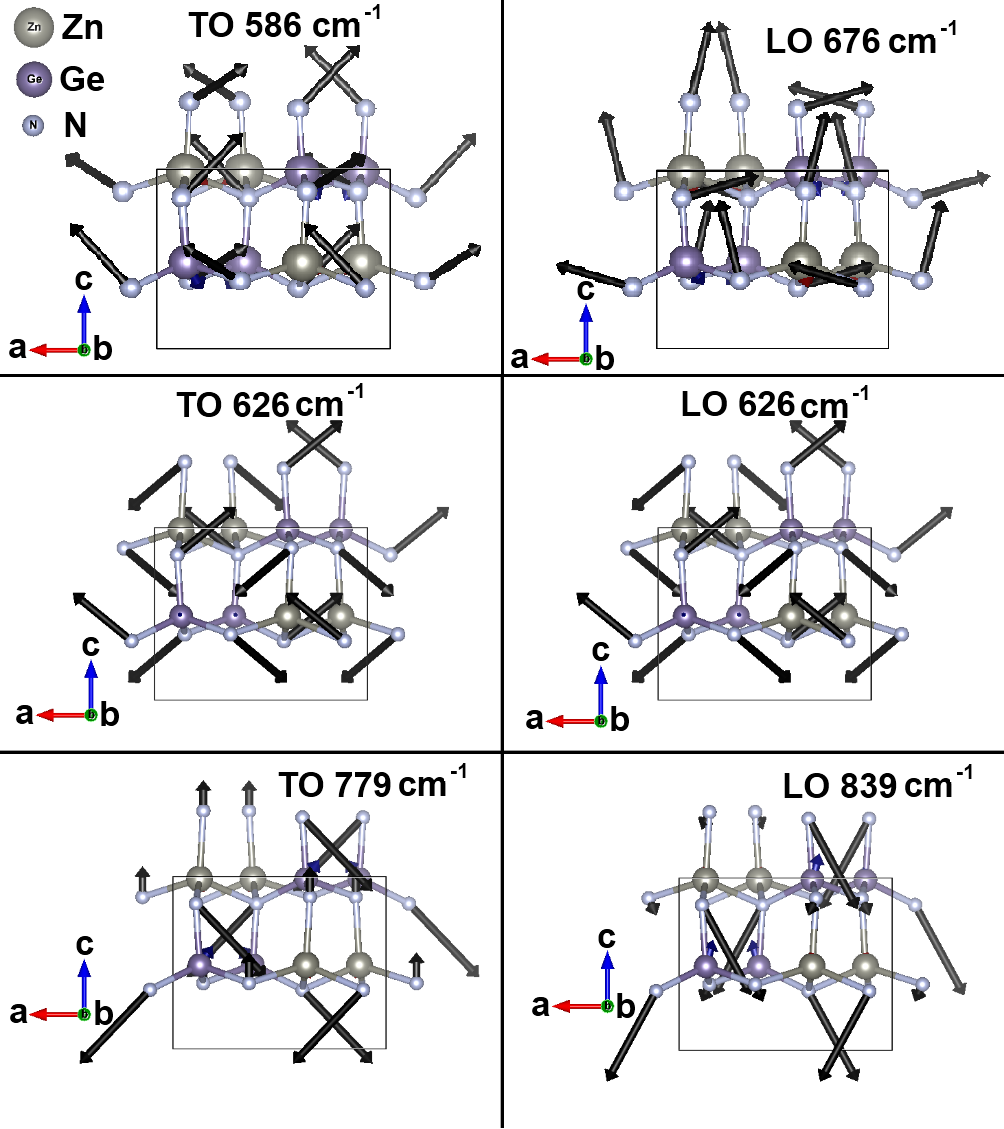}
\caption{(Color online) Normal mode patterns for three TO modes (left column) and their corresponding LO modes
 (right column).  The induced electric field is in the c-direction which is up in the figure.  The TO modes at 
586 and 779 cm$^{-1}$ induce larger electric fields and so their corresponding LO modes are altered
 accordingly.  The TO mode at 626 cm$^{-1}$ induces a small electric field so its LO mode is nearly 
unchanged.     \label{figcrossmodes}}
\end{figure}

To further emphasize the correspondence between TO and LO modes, we also 
examine the normal mode vibration patterns. 
Fig. \ref{figcrossmodes} shows the normal mode vibration patterns for some selected TO-LO mode pairs.  First, the eigenvectors
for the mode pair calculated at 626 cm$^{-1}$ (corresponding to 614.5 cm$^{-1}$ 
in the experiment) are nearly unchanged since the polarization 
and resulting electric field are very small for 
this mode, leading to a very small LO-TO frequency shift.  
The calculated TO mode at 586 cm$^{-1}$ shifts to 676 cm$^{-1}$.  In a sense this mode has crossed the mode at 626 cm$^{-1}$
 in frequency space. We can see more resemblance of the 586 cm$^{-1}$ 
eigenvector to the 676 cm$^{-1}$ mode than to the 626 cm$^{-1}$ mode. Nonetheless
we can see that some displacements in the 676 cm$^{-1}$ mode are more aligned 
with the $z$ direction, which demonstrates the effect of the electric field 
in the $z$ direction occuring in the LO mode. Notably, the 
nitrogen atom displacements in the left of the unit cell straighten out. 
The TO mode at 779 cm$^{-1}$ shifts to 839 cm$^{-1}$ due to the LO shift.  
In this case the modification to the eigenvector due to the electric field in the $z$ direction can be understood qualitatively.
The nitrogen atoms are negatively charged while the germanium and zinc atoms are positively charged.  The electric field in the
 $z$ direction then modifies the displacements of each atom type oppositely.
Also note that in all of these relatively high frequency modes the 
predominant motions occur for the nitrogen atoms.

 Two of the three LO modes that have a significant LO-TO frequency shift are not present in the measured spectra.
 The LO mode at 668 cm$^{-1}$ in the $yy$ spectrum is predicted to be large, but it is not seen in either experimental
 spectrum. Likewise, the LO mode at 839 cm$^{-1}$ is not seen in either measured spectrum.  There is a broad feature
 in both measured spectra around 536 cm$^{-1}$ that is close to the third predicted LO mode with a large LO-TO shift 
at 528 cm$^{-1}$.  However, this feature could also be associated 
with a DOS feature.  Generally, the LO modes lie close to DOS peaks 
and could be hidden by them but the DOS features in the 650-700 cm$^{-1}$ 
do not stand out above the background and the predicted LO peak  at 
839 cm$^{-1}$ lies definitely above the highest DOS peaks.

We hypothesize that these LO modes with large induced electric fields are being suppressed due to 
an overdamped plasmon coupling.  
LO phonon modes couple to the plasma through the
 electric field.  In GaN, the low carrier mobility dampens the coupled LO-plasmon mode.  In GaN doped
 with Si to a concentration of 2.5$\times$10$^{18}$ cm$^{-3}$, the LO $A_1$ mode is already broadened to the point where it cannot be 
separated from the background noise.\cite{Kozawa94}  
In Kozawa {\sl et al.} \cite{Kozawa94} 
a plasmon broadening factor of 930  cm$^{-1}$ was extracted for the highest
carrier concentration sample where the peak is still visible.
 This result corresponds to a carrier mobility of 
order 50 cm$^2$/Vs.
We may expect reasonably that a similar or even larger broadening 
and carrier concentration occurs in our ZnGeN$_2$ samples. 
The carrier concentration of the present ZnGeN$_2$ platelets could presently 
 not be measured, but ZnGeN$_2$ rods grown using a similar method and at a similar temperature were measured and a carrier concentration 
of order 10$^{19}$  cm$^{-3}$ was estimated.\cite{Dyck2016}
We have measured the spectra up to 1230 cm$^{-1}$ and did not find any 
indication of an upper LO-plasmon coupled peak in this range. 
This result 
indicates that the plasmon frequency could be even higher and outside the range measured
and indicates the carrier concentrations would be larger than $2\times10^{19}$
cm$^{-3}$. With a broadening factor $\gamma$ of order 1000-2000 cm$^{-1}$ as 
obtained by Mohajerani \etal\cite{Mohajerani} for GaN nanorods, 
our simulations of the 
Raman spectra following the approach described by Kozawa \etal\cite{Kozawa94}
indicate that the upper branch is very weak and broad 
compared to the lower branch and likely not detectable. For a carrier concentration of order $5\times10^{19}$ cm$^{-3}$ the plasmon mode would already be above 
2000 cm${-1}$. The lower branches in this case are so close to the TO mode
that they become indistinguishable from it.

With the scattering geometry used, in principle we expect to only see longitudinal $a_1$ modes.  However, because we used a high numerical aperture objective,
 there was some portion of the incident light with wavevector in directions other than the $z$ direction.  Because of this situation we might 
expect to see some intensity from TO modes.  For example, the small peak in the direction 4 spectrum at 760 cm$^{-1}$ could well be a
 trace of the strong TO $a_{1xx}$ peak predicted to be at 776 cm$^{-1}$.  Similarly, the small peak in the direction 1 spectrum at 580 cm$^{-1}$ 
could stem from the strong TO $a_{1yy}$ peak predicted at 585 cm$^{-1}$.  The third TO peak which is predicted to have a large LO-TO splitting
 is located at 491.4 cm$^{-1}$, just to the higher energy side of the large peak at 478.6 cm$^{-1}$.  In the measured spectrum in direction
 4, this TO peak might be contributing to the slight asymmetry of this large feature at 473.0 cm$^{-1}$.
These small traces of the TO Raman peaks observed further support our assignment of the crystallographic orientation of the measured platelet. As mentioned 
in the previous paragraph, the observation of these traces of the TO modes
could also be explained by the fact that they coincide with the $L-$ lower
branches of the plasmon coupled modes. Since the latter are so close to the
pure TO modes, they cannot be distinguished in the present spectra. 
In the work on GaN rods by Mohajerani \etal\cite{Mohajerani} the $L-$ modes
could be distinguished from the pure TO mode as broad peaks essentially hidden 
as subtle changes in the background and thanks to careful fitting over a range of samples with systematically changing carrier concentration. Unfortunately, 
we are not yet in a position to carry out such measurements in the present 
samples and thus our conjecture of the presence of LO-plasmon coupling is 
still indirectly based on the non-observation of LO peaks.

\section{Conclusions}
ZnGeN$_2$ was grown with a preferential c-oriented 
platelet type of crystal habit by exposing a Ge-Sn-Zn liquid to ammonia at 758$^\circ$C .
Micro-Raman measurements were carried out on these platelets for different 
orientations of the polarization directions with respect to the crystallite orientation in the plane
and with parallel or crossed incident versus scattered polarization. Combined with first-principles 
calculations these measurements allowed us to determine which were the $x$ and $y$ directions 
on the platelet and to measure the $a_2$ and $a_1$ Raman modes. Most of the predicted peak locations
were within 5 cm$^{-1}$ of the measured values, and the predicted relative intensities agreed semi-quantitatively with the measured peaks.  
The LO peaks of the three modes  that 
were predicted to have large LO-TO frequency shifts
were not detected, likely because of over-damped coupling to plasmons. This result is
consistent with the expected high unintentional n-type doping.

\acknowledgements{This work was supported by the National Science Foundation 
under grants No. DMR-1006132, DMR-1409346 (E. B and K. K.), and 
DMR-1533957 (M. H. and W. R. L.). Calculations made use of 
the High Performance Computing Resource
in the Core Facility for Advanced Research Computing at Case Western Reserve
University.}

\appendix
\section{} \label{appendix}

\begin{table}[h]
\caption{Calculated $b_1$ and $b_2$ mode frequencies and associated parameters.\label{tab1b2}}
\begin{ruledtabular}
\begin{tabular}{ccddc}
$\omega(b_{1T})$ & $\omega(b_{1L})$ & \multicolumn{1}{c}{$R_{xz}^T$} & \multicolumn{1}{c}{$R_{xz}^L$} &   $S_x$ \\
$cm^{-1}$ & $cm^{-1}$ & \multicolumn{1}{c}{$10^{-4} a_0^{-1}$} &  \multicolumn{1}{c}{$10^{-4}a_0^{-1}$} &  \multicolumn{1}{c}{a.u.=253.264 m$^2$/s$^2$} \\ \hline
\\
168.2 & 168.7 & 0.73 & 0.85 & $2.9\times10^{-6}$ \\
194.3 & 194.4 & 2.28 & 2.24 & $4.5\times10^{-7}$ \\
239.7 & 241.8 & 2.74 & 3.13 & $2.0\times10^{-5}$ \\
305.6 & 308.9 & 4.29 & 3.91 & $4.0\times10^{-5}$ \\
326.4 & 326.4 & 2.80 & 2.83 & $2.8\times10^{-7}$ \\
517.7 & 540.7 & 3.58 & 1.38 & $7.7\times10^{-4}$ \\
546.6 & 551.3 & 0.03 & 0.03 & $3.5\times10^{-3}$ \\
616.2 & 705.4 & 10.87& 15.1 & $1.3\times10^{-3}$ \\
656.3 & 656.2 & 4.06 & 3.03 & $2.8\times10^{-6}$ \\
768.9 & 821.4 & 4.00 & 19.28& $3.9\times10^{-4}$ \\
799.4 & 796.7 & 0.67 & 3.94 & $4.7\times10^{-6}$ \\ \hline 
\\
$\omega(b_{2T})$ & $\omega(b_{2L})$ & \multicolumn{1}{c}{$R_{yz}^T$} & \multicolumn{1}{c}{$R_{yz}^L$} &   $S_y$ \\ \hline
\\
133.2 & 133.4 & 0.87 & 0.80 & $9.0\times10^{-7}$ \\
167.3 & 167.3 & 2.79 & 2.80 & $9.2\times10^{-8}$ \\
208.4 & 210.3 & 3.10 & 3.41 & $1.4\times10^{-5}$ \\
270.3 & 270.6 & 1.31 & 1.47 & $2.7\times10^{-6}$ \\
341.6 & 341.7 & 1.35 & 1.27 & $1.5\times10^{-6}$ \\
478.7 & 490.6 & 1.25 & 1.80 & $6.7\times10^{-4}$ \\
508.2 & 562.2 & 0.45 & 1.26 & $7.5\times10^{-4}$ \\
599.3 & 638.8 & 1.42 & 2.70 & $3.1\times10^{-4}$\\
677.9 & 678.0 & 3.27 & 3.27 & $5.7\times10^{-7}$ \\
756.4 & 796.1 & 11.79& 9.44 & $6.7\times10^{-4}$ \\
808.2 & 842.1 & 2.47 & 19.2 & $9.9\times10^{-5}$ \\
\end{tabular}
\end{ruledtabular}
\end{table}

\bibliography{zgn,dft,abinit}

\begin{thebibliography}{18}
\expandafter\ifx\csname natexlab\endcsname\relax\def\natexlab#1{#1}\fi
\expandafter\ifx\csname bibnamefont\endcsname\relax
  \def\bibnamefont#1{#1}\fi
\expandafter\ifx\csname bibfnamefont\endcsname\relax
  \def\bibfnamefont#1{#1}\fi
\expandafter\ifx\csname citenamefont\endcsname\relax
  \def\citenamefont#1{#1}\fi
\expandafter\ifx\csname url\endcsname\relax
  \def\url#1{\texttt{#1}}\fi
\expandafter\ifx\csname urlprefix\endcsname\relax\def\urlprefix{URL }\fi
\providecommand{\bibinfo}[2]{#2}
\providecommand{\eprint}[2][]{\url{#2}}

\bibitem[{\citenamefont{Punya and Lambrecht}(2013)}]{Punya13}
\bibinfo{author}{\bibfnamefont{A.}~\bibnamefont{Punya}} \bibnamefont{and}
  \bibinfo{author}{\bibfnamefont{W.~R.~L.} \bibnamefont{Lambrecht}},
  \bibinfo{journal}{Phys. Rev. B} \textbf{\bibinfo{volume}{88}},
  \bibinfo{pages}{075302} (\bibinfo{year}{2013}),
  \urlprefix\url{http://link.aps.org/doi/10.1103/PhysRevB.88.075302}.

\bibitem[{\citenamefont{Han et~al.}(2014)\citenamefont{Han, Kash, and
  Zhao}}]{Han}
\bibinfo{author}{\bibfnamefont{L.}~\bibnamefont{Han}},
  \bibinfo{author}{\bibfnamefont{K.}~\bibnamefont{Kash}}, \bibnamefont{and}
  \bibinfo{author}{\bibfnamefont{H.}~\bibnamefont{Zhao}}, in
  \emph{\bibinfo{booktitle}{Proceedings of SPIE, Light-Emitting Diodes:
  Materials, Devices, and Applications for Solid State Lighting XVII}}
  (\bibinfo{year}{2014}), pp. \bibinfo{pages}{90030W--1--5}.

\bibitem[{\citenamefont{Blanton et~al.}(2013)\citenamefont{Blanton, He, Shan,
  and Kash}}]{Blanton13}
\bibinfo{author}{\bibfnamefont{E.}~\bibnamefont{Blanton}},
  \bibinfo{author}{\bibfnamefont{K.}~\bibnamefont{He}},
  \bibinfo{author}{\bibfnamefont{J.}~\bibnamefont{Shan}}, \bibnamefont{and}
  \bibinfo{author}{\bibfnamefont{K.}~\bibnamefont{Kash}}, in
  \emph{\bibinfo{booktitle}{Symposium E/H – Photovoltaic Technologies,
  Devices and Systems Based on Inorganic Materials, Small Organic Molecules and
  Hybrids}} (\bibinfo{year}{2013}), vol. \bibinfo{volume}{1493} of
  \emph{\bibinfo{series}{MRS Proceedings}}, pp. \bibinfo{pages}{237--242},
  \urlprefix\url{http://journals.cambridge.org/article_S1946427413002352}.

\bibitem[{\citenamefont{Quayle et~al.}(2015)\citenamefont{Quayle, Blanton,
  Punya, Junno, He, Han, Zhao, Shan, Lambrecht, and Kash}}]{Quayle15}
\bibinfo{author}{\bibfnamefont{P.~C.} \bibnamefont{Quayle}},
  \bibinfo{author}{\bibfnamefont{E.~W.} \bibnamefont{Blanton}},
  \bibinfo{author}{\bibfnamefont{A.}~\bibnamefont{Punya}},
  \bibinfo{author}{\bibfnamefont{G.~T.} \bibnamefont{Junno}},
  \bibinfo{author}{\bibfnamefont{K.}~\bibnamefont{He}},
  \bibinfo{author}{\bibfnamefont{L.}~\bibnamefont{Han}},
  \bibinfo{author}{\bibfnamefont{H.}~\bibnamefont{Zhao}},
  \bibinfo{author}{\bibfnamefont{J.}~\bibnamefont{Shan}},
  \bibinfo{author}{\bibfnamefont{W.~R.~L.} \bibnamefont{Lambrecht}},
  \bibnamefont{and} \bibinfo{author}{\bibfnamefont{K.}~\bibnamefont{Kash}},
  \bibinfo{journal}{Phys. Rev. B} \textbf{\bibinfo{volume}{91}},
  \bibinfo{pages}{205207} (\bibinfo{year}{2015}),
  \urlprefix\url{http://link.aps.org/doi/10.1103/PhysRevB.91.205207}.

\bibitem[{\citenamefont{Wintenberger et~al.}(1973)\citenamefont{Wintenberger,
  Maunaye, and Laurent}}]{Wintenberger}
\bibinfo{author}{\bibfnamefont{M.}~\bibnamefont{Wintenberger}},
  \bibinfo{author}{\bibfnamefont{M.}~\bibnamefont{Maunaye}}, \bibnamefont{and}
  \bibinfo{author}{\bibfnamefont{Y.}~\bibnamefont{Laurent}},
  \bibinfo{journal}{Mat. Res. Bull.} \textbf{\bibinfo{volume}{8}},
  \bibinfo{pages}{1049} (\bibinfo{year}{1973}).

\bibitem[{\citenamefont{Lambrecht et~al.}(2005)\citenamefont{Lambrecht,
  Alldredge, and Kim}}]{Lambrechtzgn05}
\bibinfo{author}{\bibfnamefont{W.~R.~L.} \bibnamefont{Lambrecht}},
  \bibinfo{author}{\bibfnamefont{E.}~\bibnamefont{Alldredge}},
  \bibnamefont{and} \bibinfo{author}{\bibfnamefont{K.}~\bibnamefont{Kim}},
  \bibinfo{journal}{Phys. Rev. B} \textbf{\bibinfo{volume}{72}},
  \bibinfo{eid}{155202} (\bibinfo{year}{2005}).

\bibitem[{\citenamefont{Paudel and Lambrecht}(2008)}]{Paudel08}
\bibinfo{author}{\bibfnamefont{T.~R.} \bibnamefont{Paudel}} \bibnamefont{and}
  \bibinfo{author}{\bibfnamefont{W.~R.~L.} \bibnamefont{Lambrecht}},
  \bibinfo{journal}{Phys. Rev. B} \textbf{\bibinfo{volume}{78}},
  \bibinfo{eid}{115204} (pages~\bibinfo{numpages}{12}) (\bibinfo{year}{2008}),
  \urlprefix\url{http://link.aps.org/abstract/PRB/v78/e115204}.

\bibitem[{\citenamefont{Lambrecht and Punya}(2013)}]{Lambrechtbook}
\bibinfo{author}{\bibfnamefont{W.~R.~L.} \bibnamefont{Lambrecht}}
  \bibnamefont{and} \bibinfo{author}{\bibfnamefont{A.}~\bibnamefont{Punya}}, in
  \emph{\bibinfo{booktitle}{{III-Nitride Semiconductors and their Modern
  Devices}}}, edited by \bibinfo{editor}{\bibfnamefont{B.}~\bibnamefont{Gill}}
  (\bibinfo{publisher}{Oxford University Press}, \bibinfo{year}{2013}), pp.
  \bibinfo{pages}{519--585}.

\bibitem[{\citenamefont{Viennois et~al.}(2001)\citenamefont{Viennois,
  Taliercio, Potin, Errebbahi, Gil, Charar, Haidoux, and T\'edenac}}]{Viennois}
\bibinfo{author}{\bibfnamefont{R.}~\bibnamefont{Viennois}},
  \bibinfo{author}{\bibfnamefont{T.}~\bibnamefont{Taliercio}},
  \bibinfo{author}{\bibfnamefont{V.}~\bibnamefont{Potin}},
  \bibinfo{author}{\bibfnamefont{A.}~\bibnamefont{Errebbahi}},
  \bibinfo{author}{\bibfnamefont{B.}~\bibnamefont{Gil}},
  \bibinfo{author}{\bibfnamefont{S.}~\bibnamefont{Charar}},
  \bibinfo{author}{\bibfnamefont{A.}~\bibnamefont{Haidoux}}, \bibnamefont{and}
  \bibinfo{author}{\bibfnamefont{J.-C.} \bibnamefont{T\'edenac}},
  \bibinfo{journal}{Mater. Sci. Eng. B} \textbf{\bibinfo{volume}{82}},
  \bibinfo{pages}{45} (\bibinfo{year}{2001}).

\bibitem[{\citenamefont{Peshek et~al.}(2008)\citenamefont{Peshek, Paudel, Kash,
  and Lambrecht}}]{Peshek08}
\bibinfo{author}{\bibfnamefont{T.~J.} \bibnamefont{Peshek}},
  \bibinfo{author}{\bibfnamefont{T.~R.} \bibnamefont{Paudel}},
  \bibinfo{author}{\bibfnamefont{K.}~\bibnamefont{Kash}}, \bibnamefont{and}
  \bibinfo{author}{\bibfnamefont{W.~R.~L.} \bibnamefont{Lambrecht}},
  \bibinfo{journal}{Phys. Rev. B} \textbf{\bibinfo{volume}{77}},
  \bibinfo{eid}{235213} (pages~\bibinfo{numpages}{9}) (\bibinfo{year}{2008}),
  \urlprefix\url{http://link.aps.org/abstract/PRB/v77/e235213}.

\bibitem[{\citenamefont{Gonze}(1997)}]{Gonze1}
\bibinfo{author}{\bibfnamefont{X.}~\bibnamefont{Gonze}},
  \bibinfo{journal}{Phys. Rev. B} \textbf{\bibinfo{volume}{55}},
  \bibinfo{pages}{10337} (\bibinfo{year}{1997}).

\bibitem[{\citenamefont{Gonze and Lee}(1997)}]{Gonze2}
\bibinfo{author}{\bibfnamefont{X.}~\bibnamefont{Gonze}} \bibnamefont{and}
  \bibinfo{author}{\bibfnamefont{C.}~\bibnamefont{Lee}},
  \bibinfo{journal}{Phys. Rev. B} \textbf{\bibinfo{volume}{55}},
  \bibinfo{pages}{10355} (\bibinfo{year}{1997}).

\bibitem[{\citenamefont{Gonze et~al.}(2002/11)\citenamefont{Gonze, Beuken,
  Caracas, Detraux, Fuchs, Rignanese, Sindic, Zerah, and Jollet}}]{abinit}
\bibinfo{author}{\bibfnamefont{X.}~\bibnamefont{Gonze}},
  \bibinfo{author}{\bibfnamefont{J.~M.} \bibnamefont{Beuken}},
  \bibinfo{author}{\bibfnamefont{R.}~\bibnamefont{Caracas}},
  \bibinfo{author}{\bibfnamefont{F.}~\bibnamefont{Detraux}},
  \bibinfo{author}{\bibfnamefont{M.}~\bibnamefont{Fuchs}},
  \bibinfo{author}{\bibfnamefont{G.~M.} \bibnamefont{Rignanese}},
  \bibinfo{author}{\bibfnamefont{M.}~\bibnamefont{Sindic},
  \bibfnamefont{L.and~Verstraete}},
  \bibinfo{author}{\bibfnamefont{G.}~\bibnamefont{Zerah}}, \bibnamefont{and}
  \bibinfo{author}{\bibfnamefont{F.}~\bibnamefont{Jollet}},
  \bibinfo{journal}{Computational Materials Science}
  \textbf{\bibinfo{volume}{25,3}}, \bibinfo{pages}{478}
  (\bibinfo{year}{2002/11}), \urlprefix\url{http://www.abinit.org}.

\bibitem[{\citenamefont{Perdew and Wang}(1992)}]{PerdewWang92}
\bibinfo{author}{\bibfnamefont{J.~P.} \bibnamefont{Perdew}} \bibnamefont{and}
  \bibinfo{author}{\bibfnamefont{Y.}~\bibnamefont{Wang}},
  \bibinfo{journal}{Phys. Rev. B} \textbf{\bibinfo{volume}{45}},
  \bibinfo{pages}{13244} (\bibinfo{year}{1992}),
  \urlprefix\url{http://link.aps.org/doi/10.1103/PhysRevB.45.13244}.

\bibitem[{\citenamefont{Hartwigsen et~al.}(1998)\citenamefont{Hartwigsen,
  Goedecker, and Hutter}}]{Hartwigsen98}
\bibinfo{author}{\bibfnamefont{C.}~\bibnamefont{Hartwigsen}},
  \bibinfo{author}{\bibfnamefont{S.}~\bibnamefont{Goedecker}},
  \bibnamefont{and} \bibinfo{author}{\bibfnamefont{J.}~\bibnamefont{Hutter}},
  \bibinfo{journal}{Phys. Rev. B} \textbf{\bibinfo{volume}{58}},
  \bibinfo{pages}{3641} (\bibinfo{year}{1998}),
  \urlprefix\url{http://link.aps.org/doi/10.1103/PhysRevB.58.3641}.

\bibitem[{\citenamefont{Kozawa et~al.}(1994)\citenamefont{Kozawa, Kachi, Kano,
  Taga, Hashimoto, Koide, and Manabe}}]{Kozawa94}
\bibinfo{author}{\bibfnamefont{T.}~\bibnamefont{Kozawa}},
  \bibinfo{author}{\bibfnamefont{T.}~\bibnamefont{Kachi}},
  \bibinfo{author}{\bibfnamefont{H.}~\bibnamefont{Kano}},
  \bibinfo{author}{\bibfnamefont{Y.}~\bibnamefont{Taga}},
  \bibinfo{author}{\bibfnamefont{M.}~\bibnamefont{Hashimoto}},
  \bibinfo{author}{\bibfnamefont{N.}~\bibnamefont{Koide}}, \bibnamefont{and}
  \bibinfo{author}{\bibfnamefont{K.}~\bibnamefont{Manabe}},
  \bibinfo{journal}{Journal of Applied Physics} \textbf{\bibinfo{volume}{75}},
  \bibinfo{pages}{1098} (\bibinfo{year}{1994}),
  \urlprefix\url{http://scitation.aip.org/content/aip/journal/jap/75/2/10.1063/1.356492}.

\bibitem[{\citenamefont{Dyck et~al.}(2016)\citenamefont{Dyck, Colvin, Quayle,
  Peshek, and Kash}}]{Dyck2016}
\bibinfo{author}{\bibfnamefont{J.~S.} \bibnamefont{Dyck}},
  \bibinfo{author}{\bibfnamefont{J.~R.} \bibnamefont{Colvin}},
  \bibinfo{author}{\bibfnamefont{P.~C.} \bibnamefont{Quayle}},
  \bibinfo{author}{\bibfnamefont{T.~J.} \bibnamefont{Peshek}},
  \bibnamefont{and} \bibinfo{author}{\bibfnamefont{K.}~\bibnamefont{Kash}},
  \bibinfo{journal}{Journal of Electronic Materials}
  \textbf{\bibinfo{volume}{45}}, \bibinfo{pages}{2920} (\bibinfo{year}{2016}),
  ISSN \bibinfo{issn}{1543-186X},
  \urlprefix\url{http://dx.doi.org/10.1007/s11664-015-4322-3}.

\bibitem[{\citenamefont{Mohajerani et~al.}(2016)\citenamefont{Mohajerani,
  Khachadorian, Schimpke, Nenstiel, Hartmann, Ledig, Avramescu, Strassburg,
  Hoffmann, and Waag}}]{Mohajerani}
\bibinfo{author}{\bibfnamefont{M.~S.} \bibnamefont{Mohajerani}},
  \bibinfo{author}{\bibfnamefont{S.}~\bibnamefont{Khachadorian}},
  \bibinfo{author}{\bibfnamefont{T.}~\bibnamefont{Schimpke}},
  \bibinfo{author}{\bibfnamefont{C.}~\bibnamefont{Nenstiel}},
  \bibinfo{author}{\bibfnamefont{J.}~\bibnamefont{Hartmann}},
  \bibinfo{author}{\bibfnamefont{J.}~\bibnamefont{Ledig}},
  \bibinfo{author}{\bibfnamefont{A.}~\bibnamefont{Avramescu}},
  \bibinfo{author}{\bibfnamefont{M.}~\bibnamefont{Strassburg}},
  \bibinfo{author}{\bibfnamefont{A.}~\bibnamefont{Hoffmann}}, \bibnamefont{and}
  \bibinfo{author}{\bibfnamefont{A.}~\bibnamefont{Waag}},
  \bibinfo{journal}{Applied Physics Letters} \textbf{\bibinfo{volume}{108}},
  \bibinfo{eid}{091112} (\bibinfo{year}{2016}),
  \urlprefix\url{http://scitation.aip.org/content/aip/journal/apl/108/9/10.1063/1.4943079}.

\end{thebibliography}
\end{document}